\documentclass[aps,prl,longbibliography,superscriptaddress,reprint]{revtex4-1}

\usepackage{graphicx}
\usepackage{amsmath}
\usepackage{multirow} 
\usepackage{physics} 

\begin{document}

\title{Measurement-based entanglement of noninteracting bosonic atoms}

\author{Brian J. Lester}
\altaffiliation[Current address: ]{Department of Applied Physics, Yale University, New Haven, Connecticut 06520, USA}
\email[E-mail: ]{blester@jila.colorado.edu}
\affiliation{JILA, National Institute of Standards and Technology and University of Colorado, and
Department of Physics, University of Colorado, Boulder, Colorado 80309, USA}
\author{Yiheng Lin}
\affiliation{JILA, National Institute of Standards and Technology and University of Colorado, and
Department of Physics, University of Colorado, Boulder, Colorado 80309, USA}
\author{Mark O. Brown}
\affiliation{JILA, National Institute of Standards and Technology and University of Colorado, and
Department of Physics, University of Colorado, Boulder, Colorado 80309, USA}
\author{Adam M. Kaufman}
\affiliation{JILA, National Institute of Standards and Technology and University of Colorado, and
Department of Physics, University of Colorado, Boulder, Colorado 80309, USA}
\author{Randall J. Ball}
\affiliation{JILA, National Institute of Standards and Technology and University of Colorado, and
Department of Physics, University of Colorado, Boulder, Colorado 80309, USA}
\author{Emanuel Knill}
\affiliation{National Institute of Standards and Technology, 325 Broadway, Boulder, Colorado 80305, USA}
\affiliation{Center for Theory of Quantum Matter, University of Colorado, Boulder, Colorado 80309, USA}
\author{Ana M. Rey}
\affiliation{JILA, National Institute of Standards and Technology and University of Colorado, and
Department of Physics, University of Colorado, Boulder, Colorado 80309, USA}
\affiliation{Center for Theory of Quantum Matter, University of Colorado, Boulder, Colorado 80309, USA}
\author{Cindy A. Regal}
\email[E-mail: ]{regal@colorado.edu}
\affiliation{JILA, National Institute of Standards and Technology and University of Colorado, and
Department of Physics, University of Colorado, Boulder, Colorado 80309, USA}

\begin{abstract}
We demonstrate the ability to extract a spin-entangled state of two neutral atoms via postselection based on a measurement of their spatial configuration. Typically, entangled states of neutral atoms are engineered via atom-atom interactions.  In contrast, in our work we use Hong-Ou-Mandel interference to postselect a spin-singlet state after overlapping two atoms in distinct spin states on an effective beam splitter.  We verify the presence of entanglement and determine a bound on the postselected fidelity of a spin-singlet state of $\left(0.62 \pm 0.03\right)$.  The experiment has direct analogy to creating polarization entanglement with single photons and hence demonstrates the potential to use protocols developed for photons to create complex quantum states with noninteracting atoms.
\end{abstract}

\date{\today}

\maketitle  
Neutral atoms have increasingly become a platform for understanding and characterizing entanglement in many-body systems and have the potential to become a resource for quantum information processing~\cite{islam_measuring_2015,weiss_quantum_2017}.  The advantage of neutral atoms in quantum processing is that they can be well isolated from the environment and transported spatially in close proximity with little unwanted interaction~\cite{weitenberg_quantum_2011,weiss_quantum_2017}.  However, the naturally small interactions that enable these traits have made creating entanglement between neutral atoms more challenging. To deterministically entangle the spin of individual neutral atoms, experimenters have used long-range interactions between Rydberg states~\cite{wilk_entanglement_2010,isenhower_demonstration_2010,saffman_quantum_2010} and the exchange interaction of atoms in their electronic ground state~\cite{kaufman_entangling_2015,fukuhara_spatially_2015,murmann_two_2015,anderlini_controlled_2007,trotzky_time-resolved_2008}.  It is also possible to use photons that have interacted with individual neutral atoms or ions to entangle two atomic spins~\cite{moehring_entanglement_2007, hofmann_heralded_2012,ritter_elementary_2012, casabone_heralded_2013,welte_cavity_2017}.  Many of the experiments harnessing photons to create atomic entanglement draw on the power of measurement to enable postselection or heralding, which is an increasingly common technique in atomic physics~\cite{chou_measurement-induced_2005,moehring_entanglement_2007, lettner_remote_2011,hofmann_heralded_2012,leroux_implementation_2010,bohnet_reduced_2014}.

However, controlled photon-atom interactions are not required to create entanglement via measurement.  Individual bosonic neutral atoms can themselves be interfered and detected, as in recent experiments that realize atom equivalents of the Hong-Ou-Mandel (HOM) effect~\cite{kaufman_two-particle_2014,lopes_atomic_2015}.  When neutral atoms are noninteracting, they can be used in place of photons in probabilistic entanglement schemes~\cite{ou_violation_1988,shih_new_1988,knill_scheme_2001,kok_linear_2007,pan_multiphoton_2012}. The additional advantage of choosing atoms is that many well-developed tools---addressability, single atom sources, high-efficiency single particle detection, long-lived memory---can be incorporated, which are not always accessible with photons.  In this Letter, we show that it is possible to entangle two noninteracting $^{87}$Rb atoms by postselecting on the spatial location of the atoms after their interference on a beam splitter.

\begin{figure}[t!]
	\begin{center}
		\includegraphics[width=0.9\columnwidth]{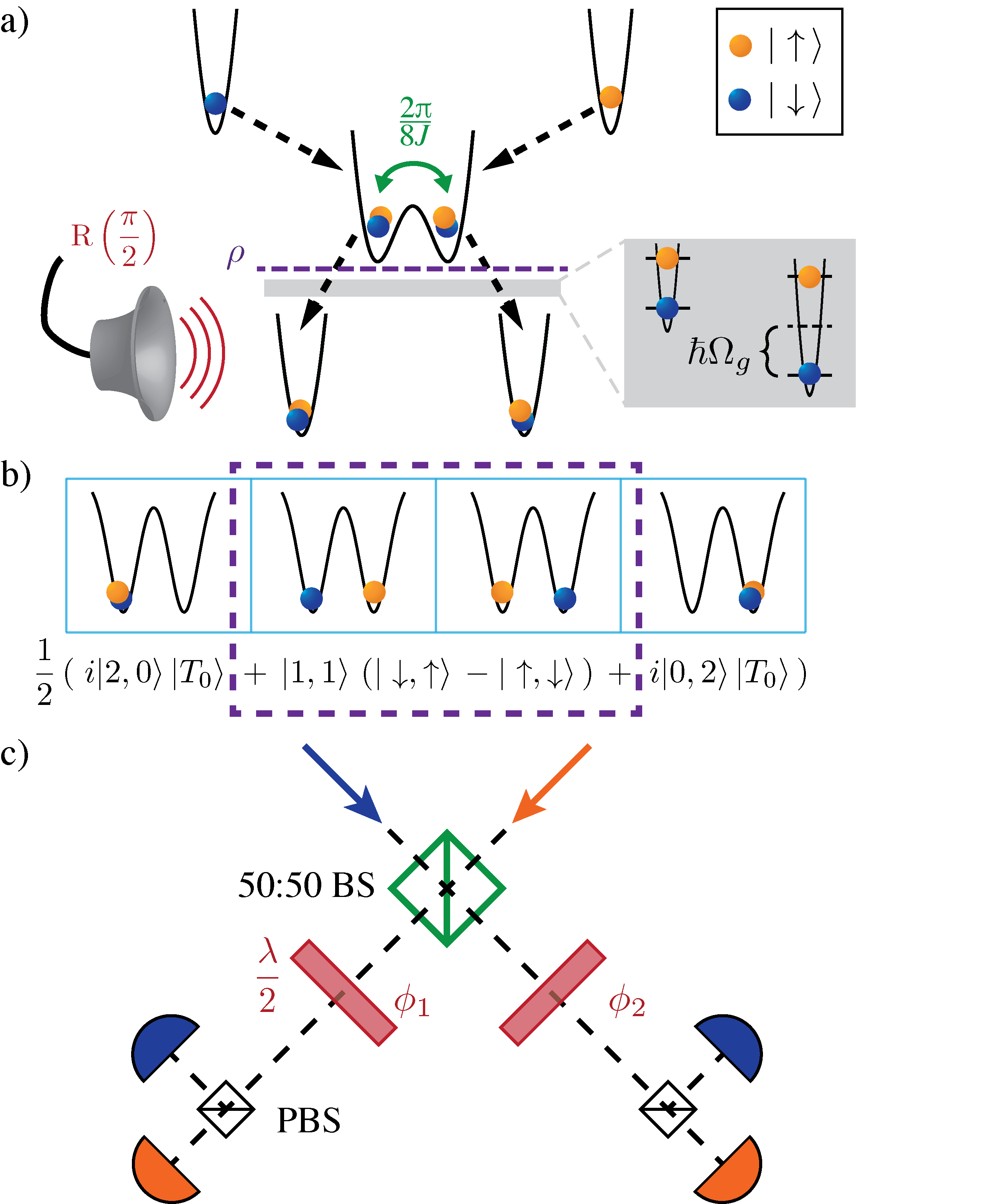}
		\caption{
			Extracting entanglement through interference and postselection. 
			a) Two bosonic atoms are prepared in orthogonal spin states and delocalized in the double-well potential to interfere the atoms via a tunnel coupling, which generates a state of interest $\rho$ (purple dashed line).  A differential spin-dependent energy shift between the wells, achieved when the wells have different depths (gray box), and a global microwave $\pi/2$ spin rotation are used to characterize entanglement in $\rho$~\cite{supplement}.
			b) The four possible measurement outcomes (blue boxes). Atoms found in separate traps, i.e.,~in the $\ket{1,1}$ spatial state, are in the maximally entangled singlet spin state $\ket{S_0}=\frac{1}{\sqrt{2}}\left(\ket{\downarrow,\uparrow}-\ket{\uparrow,\downarrow}\right)$ (purple dashed box). Atoms in the same well are in the symmetric (triplet) spin state $\ket{T_0}=\frac{1}{\sqrt{2}}\left(\ket{\downarrow,\uparrow}+\ket{\uparrow,\downarrow}\right)$. 
			c) Photon analogy. Single photons with orthogonal polarization are overlapped on a 50:50 beam splitter (BS).  When a photon is detected in each output mode, the polarization of the two photons is maximally entangled. 
			The entanglement is characterized with independent polarization rotations and a polarization-sensitive detector constructed from a polarizing beam splitter (PBS) and two photon detectors~\cite{shih_new_1988}. 
		}
		\label{MIEprocedure}
	\end{center}
\end{figure} 

We implement an effective atomic beam splitter by delocalizing atoms between two sides of a double-well potential via a resonant tunnel coupling~\cite{kaufman_two-particle_2014}~(Fig.~\ref{MIEprocedure}).  If the atoms are in the same spin state---meaning symmetric in both spin and space---and completely indistinguishable, one observes the HOM effect with atoms, where both atoms coalesce into the same well~\cite{kaufman_two-particle_2014,preiss_strongly_2015,islam_measuring_2015,lopes_atomic_2015}.  If the atoms are initialized in orthogonal spin states, they are in a superposition of the symmetric and antisymmetric spin states and, correspondingly, the symmetric and antisymmetric spatial states to preserve the total symmetry of the bosonic wavefunction.  Each of these evolve differently when combined on a 50:50 beam splitter: The symmetric portion coalesces into the same well.  The antisymmetric portion remains unchanged by the beam splitter and therefore the atoms are kept in separate wells.  Thus, by selecting cases where atoms remain in separate wells after the beam splitter, the spins will be in the maximally entangled (antisymmetric) singlet state $\ket{S_0}=\frac{1}{\sqrt{2}}\left(\ket{\downarrow,\uparrow}-\ket{\uparrow,\downarrow}\right)$, where a ket with two arrows represents the joint spin state of the atoms in well 1 and well 2, respectively.  We characterize the spin state after the effective beam splitter by performing differential spin manipulations and spin-sensitive detection.  

Our work is closely related to common experimental methods and proposals for optical photons.  In this context, it is known that interference of identical photons and strong measurement, along with phase shifting, is in principle sufficient to generate entanglement and enact quantum gates; this is the basis for linear optical quantum computing (LOQC)~\cite{knill_scheme_2001,nielsen_quantum_2000,barz_experimental_2013,spagnolo_experimental_2014,carolan_universal_2015}.   The direct optical analog to our entangling mechanism is a seminal experiment in which polarization-entangled photon states are generated by interfering two photons of orthogonal polarization on a 50:50 beam splitter and postselecting on coincident detection in the two output modes~\cite{shih_new_1988}. Figure~\ref{MIEprocedure}(c) shows the prototypical example of this experiment with linear optics, where the state is subsequently characterized using polarization-selective detectors and arbitrary polarization rotations.  By comparing the coincidence counts for particular sets of path rotation angles $\phi_1$ and $\phi_2$, Ref.~\cite{shih_new_1988} demonstrated that the correlations between the two photon polarizations violate Bell's inequalities.  Here, our goal is to demonstrate that an analogous measurement-based protocol for neutral atoms generates entangled states that could be used as a resource in quantum information processing or in construction of nontrivial many-body states.  Therefore, we verify spin entanglement of the two atoms by demonstrating that the singlet state fidelity $\mathcal{F}_{S_0}$ exceeds $1/2$.

The experiment begins by isolating two single $^{87}$Rb atoms using collisional blockade in two optical tweezers separated by $d=2.09 \; \mu$m~\cite{schlosser_sub-poissonian_2001}.  The presence (or absence) of an atom in each well of the double-well potential is recorded in an initial population image with photons collected during a period of sub-Doppler cooling. This is followed by optical pumping and three-dimensional Raman sideband cooling to initialize both atoms in the $\ket{\uparrow} \equiv \ket{F=2, \, m_F=2}$ hyperfine spin state of the $5\,S_{1/2}$ electronic orbital and in the three-dimensional motional ground state in $\left(90\pm 10\right)$\% of trials~\cite{kaufman_cooling_2012}.  We then initialize the atom in well 1 in the $\ket{\downarrow}\equiv \ket{F=1, \, m_F=1}$ state while keeping the atom in well 2 in $\ket{\uparrow}$. To achieve this, we take advantage of an added spin dependence of the trapping potential (due to the vector light shift from a small component of circular polarization in the trap light), which results in a differential energy shift $\hbar \Omega_s$ of the spin transitions when the wells have different depths. 
For our chosen trapping depths, the $\ket{\uparrow}\leftrightarrow\ket{\downarrow}$ transition for the two atoms are spectrally resolved by $\Omega_s/2\pi=153$ kHz, and we selectively rotate the spin of the atom in well 1 with a global microwave drive~\cite{supplement}.

After state preparation, the separation of the optical tweezers is adiabatically changed to bring the gaussian beam centers to $d = 900$ nm, and the trap depth is reduced to $V_0/h= \left(14.9\pm0.4\right)$ kHz per tweezer.  This realizes a tunnel coupling between well 1 and well 2 of the double-well potential with $2 \hbar J$, giving the energy difference between the ground-band spatially symmetric and antisymmetric single-particle eigenstates of the double-well~\cite{kaufman_two-particle_2014,murmann_two_2015}.  An atom initially localized in one well will be transferred to the other well with a probability that oscillates as $P_\text{tun}\left(t\right)=\frac{1}{2}\left[1-\cos\left(2 J \left(t-t_0\right)\right)\right]$.  Here $t_0$ is an offset that stems from the tunneling initialization. In the trap used for tunneling, the on-site interaction energy $U$ is small enough, with $\frac{J}{U}>3$, that the interparticle interactions do not significantly alter tunneling dynamics~\cite{supplement,kaufman_two-particle_2014,kaufman_entangling_2015,wall_effective_2015}.  After a variable period of tunneling in the double-well potential, the trap depth is diabatically increased to at least {$V_0/h=180$ kHz} to freeze tunneling dynamics.  It is at this point that the state $\rho$ [purple in Fig.~\ref{MIEprocedure}(a)] has been created; additional operations are performed to verify spin entanglement in the postselected state.

\begin{figure}[tb]
	\begin{center}
		\includegraphics[width=0.9\columnwidth]{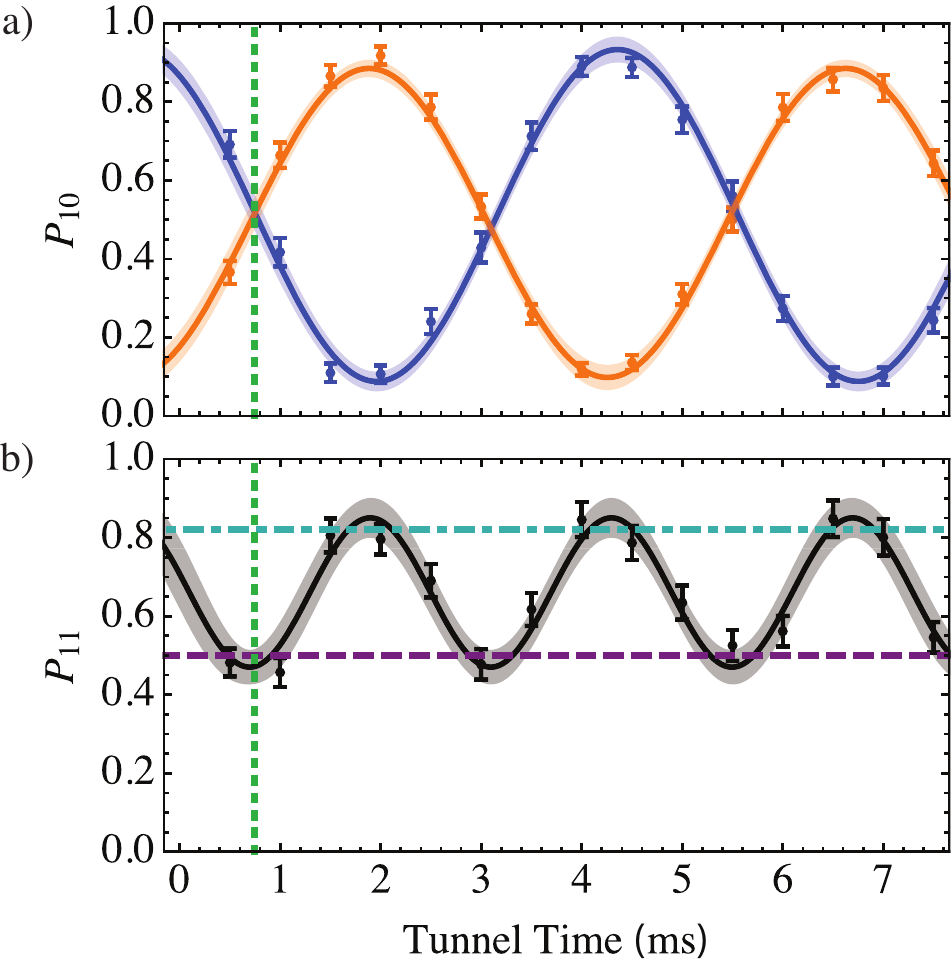}
		\caption{
			a) Measured $P_{10}$ as a function of tunneling time for a single atom initialized in $\ket{1,0}\ket{\downarrow}$ (blue) or $\ket{0,1}\ket{\uparrow}$ (orange).  The green dotted line marks the first time the 50:50 beam splitter is realized at time $t_B$.   
			b) Measured $P_{11}$ after initializing the state $\ket{1,1}\ket{\downarrow,\uparrow}$.  Also shown are the distinguishable atom limit (purple dashed line) and the maximum probability for the atoms to be in separate wells (dot-dashed cyan line) calculated from the single-atom tunneling contrast in (a). 
			All data points are plotted with error bars indicating the standard error of measurement. The fits shown are performed using a standard least-squares minimization with data points weighted by their statistical error, and the shaded regions indicate the 95\% confidence interval for the mean values predicted by the fits.
		}
		\label{parameterScans}
	\end{center}
\end{figure}

\begin{figure}[tb!]
	\begin{center}
		\includegraphics[width=0.9\columnwidth]{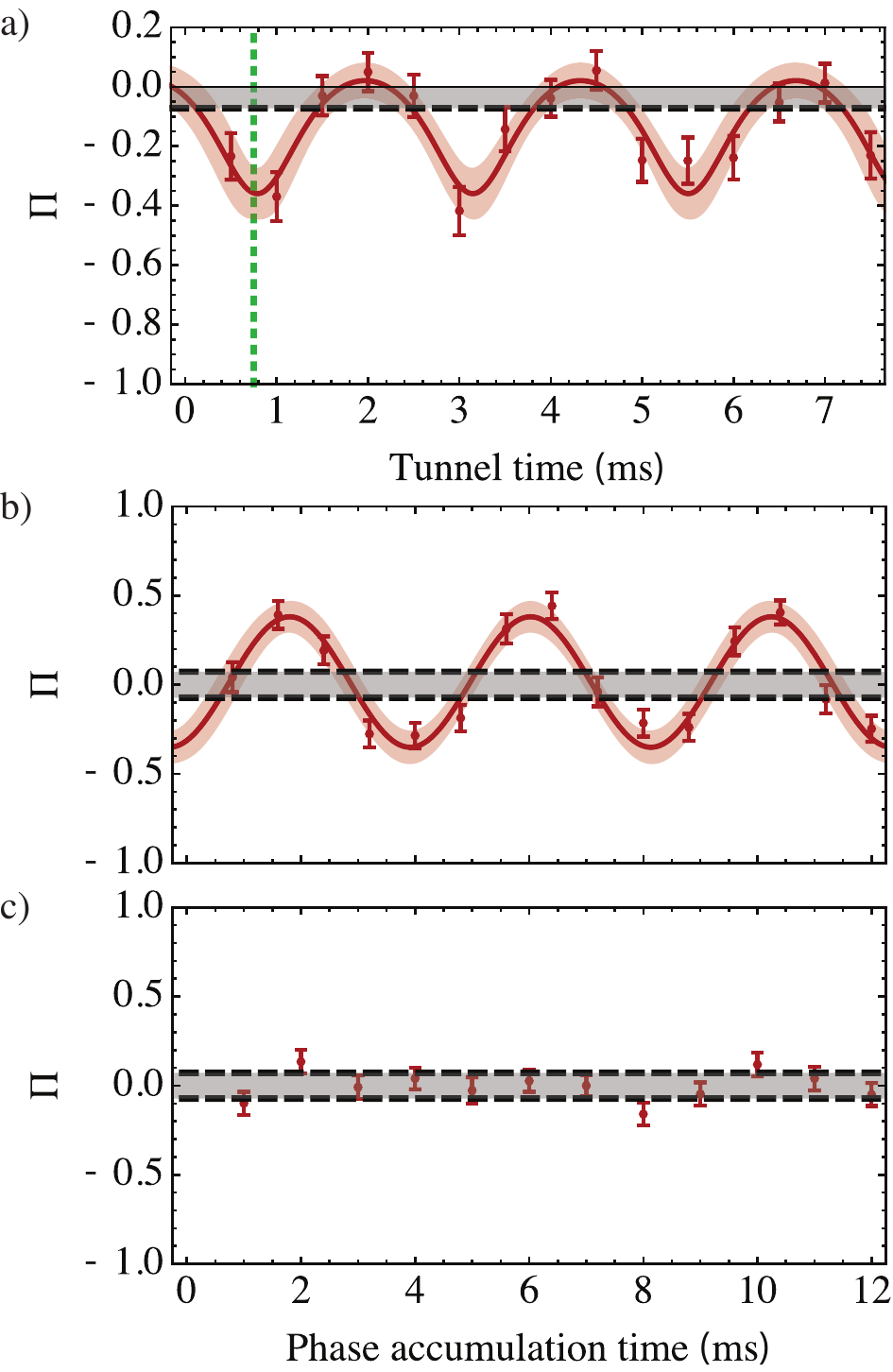}
        \caption{
			Measurements of the spin parity postselected on the $\ket{1,1}$ state after performing global microwave spin rotations. 
			a) Parity as a function of tunneling evolution time.  The minima coincide with tunneling times $t_B$ in Fig.~\ref{parameterScans} (green dotted lines). The functional form of the oscillations is expected due to postselection, as discussed in~\cite{supplement}.
			b) For a fixed tunneling evolution time of $t_B\simeq0.9$ ms, a differential phase $\Omega_g t$ is accumulated between the $\ket{\downarrow,\uparrow}$ and $\ket{\uparrow,\downarrow}$ spin states before performing the global spin rotation. For this fit, we calculate a reduced $\chi^2$ of 1.48 with 11 degrees of freedom~\cite{supplement}.
			c) The same experiment as (b), but for a fixed tunneling time of $t\ll t_B$ for which the spins remain in a separable state. 
			In all plots, the gray bars are a visual representation of  the maximum amplitude of parity oscillations $A=\pm 2 (\rho_{\uparrow \uparrow,\uparrow \uparrow} \rho_{\downarrow\downarrow,\downarrow\downarrow})^{1/2}$ that are possible for a separable density matrix, under the assumption that there are no coherences between $\ket{\uparrow,\uparrow}$ and $\ket{\downarrow,\downarrow}$.  See Fig.~2 caption for explanation of error bars.
		}
		\label{parityData}
	\end{center}
\end{figure}

Figure~\ref{parameterScans} shows the tunneling dynamics of atoms in the double-well potential to demonstrate the action of our effective atomic beam splitter.  
Note that, in all experiments presented in this manuscript, a global $\pi/2$ spin rotation $\exp\left(-i\frac{\pi}{2}\hat{S}_x\right)$, where $\hat{S}_x=\frac{1}{2}(\hat{\sigma}_x^{1}+\hat{\sigma}_x^{2})$, is applied after the beam splitter. This is the analog of setting the two wave plate angles in the photon experiment [Fig.~\ref{MIEprocedure}(c)] to $\phi_1=\phi_2=\pi/4$.  The spin rotation does not affect the population measurements, and hence is traced out in Fig.~\ref{parameterScans}.  It will be crucial, however, for inferring correlations from the projective spin measurements presented in Fig.~\ref{parityData}.  

Figure \ref{parameterScans}(a) shows the probability $P_{10}$ for a trial to end in the $\ket{1,0}$ state, where a ket $\ket{i,j}$ identifies the state with $i$ atoms in well 1 and $j$ atoms in well 2. The blue (orange) curve is for the subset of trials in which the initial population image records a single atom in well 1 (well 2).  At $t_B = \frac{2\pi}{8J}-t_0\simeq0.9$ ms, the single-particle populations cross at $P_{10}=0.5$ and the beam splitter operation is realized.  The tunneling oscillations in Fig.~\ref{parameterScans}(a), which extend to times $t\gg t_B$, are indicative of the spatial coherence of the atom. 

Figure \ref{parameterScans}(b) shows the corresponding probability $P_{11}$ for a trial to end in the $\ket{1,1}$ state for the subset of trials in which the initial population image records a single atom each in well 1 and well 2.  In the ideal situation $P_{11}=\frac{1}{2}+ \frac{1}{4}\left[1+\cos\left(4 J (t-t_0)\right)\right]$. In our measurements  we do observe that the  resulting probability for the atoms to end the tunneling sequence in separate wells oscillates at $2\pi/(4J)$, but never goes below $P_{11}=0.5$ (indicated by the purple dashed line), as expected for distinguishable bosons.  The cyan dot-dashed line represents the maximum expected $P_{11}$ based on the population dynamics measured for single-particle tunneling. Note that, for this figure only, the quantity $P_{11}$ is corrected for single particle loss; this is because imaging can not distinguish a doubly occupied well from atom loss due to the collisional blockade~\cite{schlosser_collisional_2002,kaufman_two-particle_2014,supplement}.

We now focus on the measurement-based entanglement analysis that relies on postselecting the spatial state after the beam splitter action.  The results are filtered to select only trials in which a single atom is recorded in each well in the initial population image and the $\ket{1,1}$ spatial state is measured in the final imaging sequence.  Filtering the trials on this condition not only removes trials where the final population is $\ket{0,2}$ or $\ket{2,0}$, but also removes instances where an atom is lost or the state detection protocol has an error.  For example, detection errors could come from imperfect global spin rotations or spurious large background counts.  A detailed table of the possible image outcomes for a single experimental trial, as well as their interpretation in the context of this experiment, is presented in the supplementary material~\cite{supplement}.  The postselection required for this experiment is enabled by a spin-sensitive imaging sequence that allows us to extract both spatial and spin information from each experiment trial.  Two images are taken, each of which selectively measures the population in the $\ket{\uparrow}$ spin state. In between the two images, a global $\pi$ spin rotation is applied to exchange the $\ket{\uparrow}$ and $\ket{\downarrow}$ populations~\cite{supplement,gibbons_nondestructive_2011,fuhrmanek_free-space_2011}.

In a first experiment, we reanalyze the data used for Fig.~\ref{parameterScans}(b) to study correlations in the joint spin state $\rho$ (after postselection), as a function of the tunneling time.  Specifically, we evaluate the parity of the measured spin state $\Pi = \sum_j P_j(-1)^{j}$, where $P_j$ is the probability to measure $j$ atoms in $\ket{\uparrow}$, after the global $\pi/2$ spin rotation described above.  This reanalysis gives the parity shown in Fig.~\ref{parityData}(a), which oscillates at a frequency $4J$ with the minima in parity coinciding with the minima in $P_{11}$ from Fig.~\ref{parameterScans}(b) (where a 50:50 beam splitter operation is realized).  With the knowledge that each spin is prepared in either $\ket{\uparrow}$ or $\ket{\downarrow}$ and that the tunneling dynamics conserve spin, we know a spin parity $\Pi\neq 0$ is evidence of correlations in the joint spin state $\rho$. In particular, $\Pi<0$ indicates a nonzero projection onto $\ket{S_0}$, which is an eigenstate of the global rotation and has $\Pi=-1$~\cite{supplement}. 

Next, for a fixed tunneling time $t_B$ that realizes a 50:50 beampslitter, we vary the length of time that the atoms are held in an effective magnetic field gradient between the two wells. This allows us to study the spin coherences present after the effective atomic beam splitter.  The effective gradient is provided by a spin-state-dependent relative energy shift $\hbar \Omega_g$, which is introduced with the same technique as the spin-addressing shift $\hbar \Omega_s$, but is significantly smaller with $\Omega_g/2\pi\le 0.25$ kHz~\cite{supplement}.  The spin-dependent energy shift results in the differential phase accumulation $\Omega_g t$ of the $\ket{\uparrow,\downarrow}$ state with respect to the $\ket{\downarrow,\uparrow}$ state [Fig.~\ref{MIEprocedure}(a)]. This phase accumulation ideally leads to the spin-state evolution $\ket{\Psi\left(t\right)}=\frac{1}{\sqrt{2}}\left(\ket{\downarrow,\uparrow}-e^{i\Omega_g t}\ket{\uparrow,\downarrow}\right)$, periodically rotating the singlet state to a triplet state $\ket{T_0}=\frac{1}{\sqrt{2}}\left(\ket{\downarrow,\uparrow}+\ket{\uparrow,\downarrow}\right)$ after a time $t=\pi/\Omega_g$.  Importantly, after the global $\pi/2$ spin rotation the $\ket{T_0}$ state has $+1$ parity, which results in the oscillation of the parity as a function of the accumulated differential phase $\Omega_g t$, as seen in Fig.~\ref{parityData}(b). We fit the oscillation of the measured parity as $\Pi\left(t\right)= C_\Pi \cos\left(\Omega_g t + \theta_0\right)+p_0$, which gives $C_\Pi=-\left(0.36\pm0.03\right)$, $\Omega_g/2\pi=\left(237\pm4\right)$ Hz, consistent with the expectation for $\Omega_g$~\cite{supplement}. These parity oscillations will have $C_\Pi = -1$ for a perfect singlet state. The offsets in phase and parity are $\theta_0=(0.46 \pm 0.19)$ and $p_0=(0.015\pm0.025)$, respectively.  We perform a nonparametric bootstrap analysis to check the consistency of the analysis~\cite{supplement}.  For comparison, we perform the same set of rotations when the tunneling time is short compared to $t_B$, such that the spin state is primarily $\ket{\downarrow,\uparrow}$ and observe the reduced parity signal shown in Fig.~\ref{parityData}(c).  

After perfect state preparation and an ideal beam splitter operation, the spin state postselected on the atom location $\ket{1,1}$ will be the maximally entangled spin-singlet state $\ket{S_0}$.  The singlet state fidelity is a standard entanglement witness; a fidelity $\mathcal{F}_{S_0}$ exceeding $1/2$ is sufficient to both verify entanglement and, given many copies of the same state, to distill arbitrarily good singlet states~\cite{horodecki_quantum_2009,bennet_purification_1996}. 
Here, the fidelity is given by $\mathcal{F}_{S_0}= \bra{S_0} \rho \ket{S_0}$, where $\rho$ is the $4 \times 4$ density matrix of the postselected joint spin state (after the beam splitter, but before the spin manipulations), which becomes
	\begin{align}
	\label{fidelity}
		\mathcal{F}_{S_0}
			&= \frac{1}{2}\left(\rho_{\uparrow\downarrow,\uparrow\downarrow} +  \rho_{\downarrow\uparrow,\downarrow\uparrow}\right) - \text{Re} \left(\rho_{\uparrow\downarrow,\downarrow\uparrow}\right),
	\end{align} 
with $\rho_{i,j}$ indicating the density matrix elements for the possible spin configurations $i$ and $j$. 

These density matrix elements can be bounded by combining the measurements described above with external characterizations of our state preparation and single-spin coherence~\cite{supplement}.  Specifically, a lower bound on the first two terms in Eqn.~\ref{fidelity} is given by $\rho_{\uparrow\downarrow,\uparrow\downarrow} + \rho_{\downarrow\uparrow,\downarrow\uparrow}\ge\left(0.870\pm0.018\right)$, where the spin populations are determined by a separate measurement of the spin population after the initial state preparation.  This represents a lower bound because the HOM effect results in atoms with aligned spins contributing relatively less due to postselection on $\ket{1,1}$.  The parity oscillation contrast measured in Fig.~\ref{parityData}(b) is ideally a direct measurement of the third term in Eqn.~\ref{fidelity}, i.e.,  $C_\Pi=2 \,\text{Re}\left(\rho_{\uparrow\downarrow,\downarrow\uparrow}\right)$.  This equality remains valid by assuming, as justified in~\cite{supplement}, that no coherences exist between the $\ket{\uparrow,\uparrow}$ and $\ket{\downarrow,\downarrow}$ states (before the two spin manipulations). With these measurements, we calculate a postselected singlet state fidelity of $\mathcal{F}_{S_0}\ge\left(0.62\pm0.03\right)$.

We note that the finite contrast of the parity oscillations, while large enough to verify the presence of entanglement, indicates imperfections in the spin preparation and tunneling initialization~\cite{kaufman_two-particle_2014}.  The measured parity contrast is consistent with expectations from separate measurements of the atomic HOM interference contrast, and can be improved with higher-fidelity state preparation and tunneling procedures~\cite{supplement}.

Through the interference of neutral atoms, we have demonstrated that postselection on the spatial configuration of atoms can be used to isolate spin-entangled states.  Measurement-based schemes can be extended to entanglement of larger and more complex systems by determining the success of the entire operation based on the final population distribution~\cite{kok_linear_2007,pan_multiphoton_2012,carolan_universal_2015}.  Alternatively, the presence of entanglement can be heralded through measurement of a subsystem, which makes the desired state available for subsequent steps in a larger protocol~\cite{pittman_heralded_2003}.  The simplest way to envision this possibility is to introduce a strong on-site interaction, such as through photoassociation, at the end of the protocol to expel spatial states with two atoms on a well, while leaving states with one atom per well unaffected~\cite{stwalley_photoassociation_1999}.  It is also possible to herald the presence of an entangled state without adding interactions by introducing ancilla atoms and wells in analogy to demonstrations with photons~\cite{zhang_demonstration_2008}.  

\begin{acknowledgments}
We thank Michael Foss-Feig, Leonid Isaev, and Tobias Thiele.  We acknowledge funding from NSF Grant No.~PHYS 1734006, ONR Grant No.~N00014-17-1-2245, AFOSR-MURI Grant No.~FA9550-16-1-0323, a Cottrell Scholar award, and the Packard Foundation.  A.M.R.~acknowledges funding from AFOSR FA9550-13-1-0086, AFOSR-MURI Advanced Quantum Materials, NIST and DARPA W911NF-16-1-0576 through ARO.  M.O.B.~acknowledges support from an NDSEG fellowship. 
\end{acknowledgments}


%


\clearpage

\onecolumngrid
\begin{center}
\textbf{\large Supplementary Materials for: \\ 
		``Measurement-based entanglement of noninteracting bosonic atoms''
		}

\end{center}
\twocolumngrid

\setcounter{equation}{0}
\setcounter{figure}{0}
\setcounter{table}{0}
\setcounter{page}{1}
\makeatletter
\renewcommand{\theequation}{S\arabic{equation}}
\renewcommand{\thefigure}{S\arabic{figure}}
\renewcommand{\thetable}{S\Roman{table}}

\section{Atom preparation in optical tweezers}

\subsection{Optical tweezer and tunneling parameters}

The parameters of the optical tweezer traps used in this experiment are similar to those described in Ref.~\cite{kaufman_entangling_2015}, where an optical tweezer is formed by focusing trap light to a $1/e^2$ spot radius of $w\simeq 0.7\;\mu$m.  In this work, we use temporally incoherent ($\Delta \lambda \simeq 3$ nm), far-off-resonant ($\lambda_\text{peak}=852$ nm) light to form the trap, which reduces the effect of slowly-varying interference from stray reflections in the experiment.  

For implementing beam splitter operations, two wells are brought together and partially overlapped to form a double-well potential with a tunable tunnel coupling between the two minima.  To ensure that the two-atom dynamics are representative of noninteracting bosons, we compare the tunneling energy scale $J$ to the on-site interaction strength $U=\frac{4\pi\hbar^2a_s}{m}\int{d^3\vec{r}|\psi_{g}(\vec{r})|^4}$~\cite{kaufman_two-particle_2014,kaufman_entangling_2015,wall_effective_2015}, where $a_s$ is the $s-$wave scattering length, $m$ is the mass of $^{\rm 87}\rm Rb$, and $\psi_{g}$ is the wavefunction for the ground motional state. To estimate the parameters during tunneling, we calculate $U/h=28.4^{+8.8}_{-4.5}$ Hz~\cite{wall_effective_2015}, using the trap parameters $w=\left(700\pm100\right)$ nm, atom spacing $d=\left(901\pm13\right)$ nm, individual trap depth $V_0/h=\left(14.9\pm4\right)$ kHz. 

\subsection{Site-selective spin manipulation}

After ground state cooling, we optically pump both $^{\rm 87}$Rb atoms to the $\ket{\uparrow}$ state. During optical pumping, an external magnetic field of approximately 3 G is applied, which sets the quantization axis and splits the degeneracy of the hyperfine levels in each spin manifold.  As described in the main text, the final step of our state preparation procedure initializes the atom in well 1 in the $5\,\text{S}_{1/2}\,F=1, \, m_F=1$ spin state $\ket{\downarrow}$ while keeping the atom in well 2 in the $\ket{\uparrow}$ state. To achieve this, we take advantage of an added spin-dependence of the trapping potential that is realized by introducing a small circular component to the polarization of the trap light. This provides a differential shift of the spin states when the two wells are different depths. Parameters are chosen to generate a relative shift of the spin transitions of the two atoms of $\Omega_s/2\pi=153$ kHz.  We then apply a global microwave field that is resonant with only the atom in well 1 such that, by performing a $\pi$-rotation, we create the initial state $\ket{\downarrow,\uparrow}$. 

The spectral resolution of the spin transitions between the two traps is achieved using the following polarizations and trap configurations:  The trap polarization is approximately represented by the Jones' vector $\mathcal{N} (1.00, -0.02 + 0.05i)$, where $\mathcal{N}$ is a normalization factor. The tweezer depths are ramped to approximately $V_0^{1}/h = 1.58$ MHz for well 1 and $V_0^2/h=12.7$ MHz for well 2. This imbalanced configuration creates an effective magnetic field gradient of approximately 0.073 G between the traps (which are spaced by 2.09 $\mu$m for site-selective spin rotation) and is directed along the axis of the optical tweezer traps, which is sufficient to spectrally resolve the spin transitions between the two traps.

Importantly we can effectively turn on and off $\Omega_s$ by aligning or anti-aligning an external quantization field that is much larger than the effective magnetic field gradient.  If the external field is parallel (orthogonal) to the effective field gradient (i.e., the axis of the trap beams), the differential magnitude of the total field, is maximized (minimized) to give a 153 kHz (2 kHz) shift of the spin-resonance. For the microwave spin rotation Rabi frequency $\omega_{\rm mw}=\pi/t_{\pi}\approx2\pi\times36$ kHz used here, these shifts are sufficient to turn on and off the individual addressing.

Further, we utilize a smaller effective magnetic field gradient to achieve the shift $\Omega_g$ for the parity analysis in Figs.~3(b) and~3(c).  For this experiment, we set the trap depths to be approximately 153 kHz for well 1 and 173 kHz for well 2, and align the quantization magnetic field with the trap axis. In this configuration, we realize a differential spin-shift of $\Omega_g/2\pi \simeq264$ Hz, as shown by the oscillations in Fig.~3(b). 

While the introduction of a circular component to the tweezer light is useful for state preparation and readout, it also has the potential to introduce unwanted spin resonance shifts that could dephase the desired entangled state.  In previous work studying spin-exchange entanglement, the fidelity was extremely sensitive to imperfect polarization of the trapping light~\cite{kaufman_entangling_2015}.  The origin of this susceptibility is the fact that for on-site spin-exchange the atoms have to be overlapped and then separated using techniques that require significant imbalance in the trap depths.  In contrast, in this work the tunneling occurs when the two wells are approximately balanced in intensity. Thus, the $\ket{\uparrow,\downarrow}$ and $\ket{\downarrow, \uparrow}$ states remain largely degenerate.  However, the wells used in the experiment are oriented vertically, and hence there is a small effect from gravity that results in the two wells having estimated depths of $V_0^1/h=15.8$ kHz and $V_0^2/h=14.0$ kHz, giving a differential spin resonance frequency of 24 Hz when the magnetic field is aligned to maximize the differential shift (for applying a magnetic field gradient to analyze the entangled state), and just 1.6 Hz when magnetic field is orthogonal to minimize the shift (for tunneling without a gradient). In both cases this effect is small compared to the tunneling energy scale $J/h=$104 Hz. Further, we do not observe a difference in the parity contrast in Fig.~3(a) and (b), which suggests we are not limited by the residual effective gradient.

\section{Detection and state characterization protocols}

\subsection{Spin-selective imaging}

We detect the presence of atoms and their spin state by applying a laser beam with $\sigma^+$ polarization along the direction of the external quantization magnetic field. This beam is resonant with the cycling transition between the $\ket{\uparrow}$ state to the $F^\prime=3$, $m_{\rm F}^\prime=3$ state of the $5 P_{3/2}$ electronic orbital,  and we collect the atom fluorescence on a EM-CCD camera. To distinguish atom loss from populations in the $\ket{\downarrow}$ state, we subsequently apply a global microwave $\pi$-rotation driving the $\ket{\uparrow}\leftrightarrow\ket{\downarrow}$ transition, and repeat the above fluorescence measurement to obtain information related to the populations in the $\ket{\downarrow}$ state. Thus we have two consecutive images of the atom, from which we extract populations in the $\ket{\uparrow}$ and $\ket{\downarrow}$ states during the same experimental trial. 

It was observed that driving resonant transitions while the trap light was on resulted in the collection of a smaller number of photons than expected. As suggested in Ref.~\cite{hutzler_eliminating_2017}, this problem is likely due to anti-trapping of atoms in the excited states due to light shifts from the trap light, which results in faster heating and a reduced averaged trap depth.   This problem was mitigated by rapidly turning on and off the trap light and the resonant light out-of-phase with each other at a rate of approximately 800 kHz. We use a duty cycle in which the resonant beam is on 40\% of the time and the trap beam on 50\% of the time, in order to ensure the resonant light is never on at the same time as the trap light. 

During a 3 ms period of imaging a signal corresponding to approximately 45 photons is detected on the camera if the atom is in $\ket{\uparrow}$ state, in comparison to approximately 2 photons when there is no atom or the atom  is in the $\ket{\downarrow}$ state. Due to technical issues in the detection, there is a 5\% overlap in histogram for the events with and without fluorescence for a single picture.  However, by utilizing both images we filter out trials where the fluorescence for the two images gives an invalid or inconsistent result, as indicated in Table \ref{table:twoQubitDetection} and Table \ref{table:twoQubitDetectionPostselectedAway}. The signal for a trial passes the filtering only if the fluorescence is above the threshold in only one of the two images for each well. Then, a spin-detection error on a single well goes undetected by the filtering when both images give erroneous result, with a probability of $(5\%)^2=2.5\times10^{-3}$, which is negligible for this experiment.

In previous work \cite{kaufman_entangling_2015}, we applied a resonant beam to remove atoms in the $\ket{\uparrow}$ state and imaged with a spin-insensitive polarization-gradient-cooling (PGC) technique. While the PGC imaging gives very good signal-to-noise because atoms are kept cold during photon scattering, this technique cannot distinguish atom loss during the experiment from an atom in the $\ket{\uparrow}$ state. Thus, in the PGC imaging, separate measurements are needed to estimate the loss rate in order to deduce both the $\ket{\uparrow}$ and $\ket{\downarrow}$ state populations. 

\begin{table*}[t]
  	\begin{tabular}{*{5}{|c}|}
		\hline
		Population image  & \multicolumn{4}{c|}{01}  \\ \hline
    	$\ket{\uparrow}$ image     & 01 & 00     & 10 & 00     \\ \hline
    	$\ket{\downarrow}$ image    & 00 & 01     & 00 & 10     \\ \hline 
    	Interpretation  &  $\;\left(\ket{0,1}\rightarrow\ket{0,1}\right)\otimes\ket{\uparrow}\;$ & $\;\left(\ket{0,1}\rightarrow\ket{0,1}\right)\otimes\ket{\downarrow}\;$ & $\;\left(\ket{0,1}\rightarrow\ket{1,0}\right)\otimes\ket{\uparrow}\;$ & $\;\left(\ket{0,1}\rightarrow\ket{1,0}\right)\otimes\ket{\downarrow}\;$ \\
          &  Not tunneled & Not tunneled & Tunneled &Tunneled \\ \hline \hline
    	Population image & \multicolumn{4}{c|}{10} \\ \hline 
    	$\ket{\uparrow}$ image     & 10 & 00     & 01 & 00 \\ \hline
    	$\ket{\downarrow}$ image      & 00 & 10     & 00 & 01    \\ \hline
    	Interpretation  &  $\;\left(\ket{1,0}\rightarrow\ket{1,0}\right)\otimes\ket{\uparrow}\;$ & $\;\left(\ket{1,0}\rightarrow\ket{1,0}\right)\otimes\ket{\downarrow}\;$ & $\;\left(\ket{1,0}\rightarrow\ket{0,1}\right)\otimes\ket{\uparrow}\;$ & $\;\left(\ket{1,0}\rightarrow\ket{0,1}\right)\otimes\ket{\downarrow}\;$ \\
          &  Not tunneled & Not tunneled & Tunneled &Tunneled \\\hline \hline
    	Population image &  \multicolumn{4}{c|}{11} \\ \hline
    	$\ket{\uparrow}$ image     & 11 & 00  & 01 & 10     \\ \hline
    	$\ket{\downarrow}$ image    & 00 & 11  & 10 & 01   \\ \hline
    	Interpretation & $\ket{\uparrow,\uparrow}$ & $ \ket{\downarrow,\downarrow}$ & $ \ket{\downarrow,\uparrow}$ & $\ket{\uparrow,\downarrow}$  \\
          &  $\Pi=+1$ & $\Pi=+1$ & $\Pi=-1$ & $\Pi=-1$ \\ \hline
  	\end{tabular}
	\caption{
    Summary of relevant detection pathways and their interpretations. The
labels 00, 01, 10, 11 indicate the state recorded on the EM-CCD camera for each of the two pixels; a $1$ indicates ``bright'' detection events, where the number of counts on a single pixel exceeds a measured threshold, while a zero indicates ``dark'' detection events. The first two sections of the table enumerate the physical outcomes when only one atom is recorded in the initial population image. The third section shows the interpretations when one atoms is initially loaded in each well and one atom ends in each well; in these experiments, the final spin state (and, for Fig.~3, the parity of the state) is of interest. For all three sections, we include only the detection pathways that are included in the analyses shown in Figs.~2 and~3.
	}
	\label{table:twoQubitDetection}
\end{table*}

Tables~\ref{table:twoQubitDetection} and~\ref{table:twoQubitDetectionPostselectedAway} demonstrate the pathways for data analysis used in Figs.~2 and 3 of the main text.  The first line is the configuration achieved with PGC imaging that measures population before the experiment begins.  The second two lines indicate the result of the two spin images with a global microwave $\pi$-rotation in between.  The last line indicates the interpretation of the outcome of the experimental run based upon this information.  Table~\ref{table:twoQubitDetection} highlights the trials of interest for single and two-particle tunneling, and Table~\ref{table:twoQubitDetectionPostselectedAway} details the trials that do not result in a final $\ket{1,1}$ states or contain a detection error.  We also tabulate the absolute number of counts received in each case.

\subsection{Imaging with collisional blockade}
The collisional blockade is an effect where two atoms in close proximity collectively scatter an off-resonant photon via an attractive excited-state molecular potential~\cite{schlosser_collisional_2002,kaufman_two-particle_2014}. In this process, both atoms gain a large amount of kinetic energy, resulting in loss of one or both atoms from the optical trapping potential.  While this can in principle be controlled in certain situations~\cite{weiner_experiments_1999,grunzweig_near-deterministic_2010,lester_rapid_2015}, these are not the same conditions that are ideal for imaging in this experiment.   

This effect is specifically relevant to the data presented in Fig.~2(b), where to calculate $P_{11}$ we need to calculate the ratio of events where two intially loaded atoms end in separate wells to events where both atoms end in a single well. However, due to the collisional blockade, atoms ending in the same well provides the same experimental signature as single-atom loss during the experiment. Thus, to find the oscillation amplitude of $P_{11}$ that is comparable to those measured in all other data presented, we must correct the raw measured data $P_{11}^{\star}$ by the single-particle survival probabilities for each atom, $P_{\text{survive}}^{1}$ and $P_{\text{survive}}^2$, which are extracted from the same measurements used to generate Fig.~2(a).  Here, $P_{11}^\star$ represents the probability to measure the spatial state $\ket{1,1}$ after the initial population image records an atom in each well.  Therefore, the data presented in Fig.~2(b) is calculated via

\begin{align}
\label{P11_lossCorrection}
	P_{11} &= \frac{P_{11}^\star}{P_{\text{survive}}^1 P_{\text{survive}}^2}.
\end{align}

\begin{table*}[t]
  	\begin{tabular}{*{20}{|c}|}
      \hline
          Population image  & 00 & \multicolumn{6}{c|}{01 or 10}   & \multicolumn{12}{c|}{11}  \\ \hline
          $\ket{\uparrow}$ image & any & \multicolumn{2}{c|}{00}  &  \multicolumn{3}{c|}{01 or 10} & 11           & \multicolumn{3}{c|}{00}    & \multicolumn{3}{c|}{01}   & \multicolumn{3}{c|}{10}   & \multicolumn{3}{c|}{11} \\ \hline
          $\ket{\downarrow}$ image  & any    & 00 & 11                  & 01 & 10 & 11                   & any           & 00 & 01 & 10               & 00 & 01 & 11              & 00 & 10 & 11              & 01 & 10 & 11\\ \hline
          events & 2053    & 890 & 3 & 4 & 7 & 0 & 1           & 905 & 395 & 427 & 570 & 65 & 2 & 612 & 67 & 1 & 3 & 0 & 0 \\ \hline          
         interpretation  & no atoms loaded           & \multicolumn{6}{c|}{atom lost or detection error}         & \multicolumn{12}{c|}{\begin{tabular}{@{}c@{}} $P_{\rm other,11} = $ atom lost or overlapped \\or detection error \end{tabular}}\\ \hline
  	\end{tabular}
    
 	\vspace{4mm}
  	
    \begin{tabular}{*{4}{|c}|}
         \hline 
         case & events & subcase  & events \\ \hline
         no atoms loaded & 2053   &     &    \\
         \hline
          \multirow{2}{*}{only atom\#1 loaded} & \multirow{2}{*}{3108}   &  detect single atom after tunneling   & 2698   \\ 
          \cline{3-4}
          &   &  atom loss or detection error   & 410   \\ 
           \hline
            \multirow{2}{*}{only atom\#2 loaded} & \multirow{2}{*}{3882}   &  detect single atom after tunneling   & 3387   \\ 
          \cline{3-4}
           &   &  atom loss or detection error   & 495   \\ 
           \hline
           \multirow{2}{*}{both atoms loaded} & \multirow{2}{*}{5957}   &  \textbf{detect one atom in each well after tunneling}   & \textbf{2910}   \\ 
          \cline{3-4}
           &   &  atom loss or overlap or detection error   & 3047  \\
           \hline
  	\end{tabular}
	\caption{The upper table represents the detection protocol for events that are removed because of atom loss or detection errors. Here ``any" means the interpretation is the same regardless of the image configuration. When the total bright pixel number combining the two spin images is different from the initial population image, the cause may be due to atom loss, errors in the detection, or two atoms overlapped in the same well. We also show the corresponding event numbers for various cases for the data shown in Fig.~3(b).  The lower table are more statistics for data of Fig.~3(b), showing the cases that lead to one atom in each well after tunneling, from which we further analyze the spin configurations according to Table \ref{table:twoQubitDetection} to give Fig.~3. The entire Fig.~3(b) data set contains 15,000 events in total. 
    }
    \label{table:twoQubitDetectionPostselectedAway}
\end{table*}

\section{Data Analysis}

\subsection{Functional form of fit to parity during tunneling}
In Fig.~3(a), we show the oscillation of the measured parity as a function of tunneling time. The shape of the fit is expected because we are plotting the parity only of atoms that remain in separate wells at the end of the tunneling period, which is itself a quantity that is oscillating sinusoidally [as seen in Fig.~2(b)].  The functional form of this fit is

	\begin{align}
    \label{fitEqn1}
    	f\left(x\right) &=  A \;\frac{ \left(1-\cos\left(2\pi f_\text{tun} t + \phi_\text{tun}\right)\right) }{4 - \left(1-\cos \left(2\pi f_\text{tun} t + \phi_\text{tun}\right)\right)} + \text{offset}.
    \end{align}

The parity can be thought of as a measurement of the difference in the population of symmetric vs antisymmetric spin states (before the global $\pi/2$ spin rotation), divided by the total population in the postselected spatial state. Therefore, in Fig.~3(a) we fit to the function given by \ref{fitEqn1}, which has a numerator oscillating sinusoidally from zero to a fitted maximum that represents the difference in the antisymmetric and symmetric populations. The denominator oscillates from a maximum value down of one to a minimum value of half and represents the total population in separate wells. The two amplitudes of these oscillations are combined into a total oscillation amplitude $A$ and then an offset is added as a sanity check (that is consistent with 0 in all fitted results).

\subsection{Fidelity}
After allowing two atoms to tunnel for a period of time and postselecting on the $\ket{1,1}$ state, we can generally express the spin state of the atoms as a density matrix $\rho=\sum_{i,j}\rho_{i,j}\ket{i}\bra{j}$, where $i$ and $j$ are each one of $\{\uparrow\uparrow,\uparrow\downarrow,\downarrow\uparrow,\downarrow\downarrow\}$, and $\rho_{i,j}$ are the elements of the density matrix. We are interested in the singlet state fidelity $\mathcal{F}_{S_0}=\bra{S_0}\rho\ket{S_0}=\frac{1}{2}(\rho_{\uparrow\downarrow,\uparrow\downarrow}+\rho_{\downarrow\uparrow,\downarrow\uparrow})-{\rm Re}(\rho_{\downarrow\uparrow,\uparrow\downarrow})$, which reflects the overlap of the targeted singlet state with the experimental state we produce.  The diagonal components $\rho_{i,i}$ can be measured directly with the spin-selective imaging procedure, detailed in the previous section.  To measure ${\rm Re}(\rho_{\downarrow\uparrow,\uparrow\downarrow})$, we apply a microwave $\pi/2$-rotation and perform spin-selective measurement~\cite{casabone_heralded_2013}.  The population of state $\ket{i}$ after the $\pi/2$ spin rotation is denoted as $\rho_{i,i}^{\pi/2}$.  From these measurements, we construct the parity $\Pi=\rho_{\uparrow\uparrow,\uparrow\uparrow}^{\pi/2}+\rho_{\downarrow\downarrow,\downarrow\downarrow}^{\pi/2}-\rho_{\uparrow\downarrow,\uparrow\downarrow}^{\pi/2}-\rho_{\downarrow\uparrow,\downarrow\uparrow}^{\pi/2}$. If coherences between the $\ket{\uparrow,\uparrow}$ and $\ket{\downarrow,\downarrow}$ states can be neglected, the measured parity is equal to $2\, {\rm Re}(\rho_{\uparrow\downarrow,\downarrow\uparrow})$. Under this assumption, which is justified in the main text, we can estimate the fidelity of the generated state to a singlet spin state as $\mathcal{F}_{S_0}=\frac{1}{2}(\rho_{\uparrow\downarrow,\uparrow\downarrow}+\rho_{\downarrow\uparrow,\downarrow\uparrow}-\Pi)$.  As described in the main text, we measure the initial spin preparation giving the bound $\rho_{\uparrow\downarrow,\uparrow\downarrow}+\rho_{\downarrow\uparrow,\downarrow\uparrow}\ge\left(0.870\pm0.018\right)$ and from the contrast of the parity oscillations in Fig.~3(b) we find $\Pi=C_\Pi=\left(-0.36\pm0.03\right)$, giving the quoted value of the fidelity $\mathcal{F}_{S_0}\ge\left(0.62\pm0.03\right)$.

The unlikely presence of coherences between the $\ket{\uparrow,\uparrow}$ and $\ket{\downarrow,\downarrow}$ spin states would result in parity oscillations that could, in principle, mimic the data shown in the main text Figure 3(b). Two independent observations indicate there these unintentional coherences are not present:  First, the measured single-spin dephasing time is $T_2^\star\le0.2$ ms, while the length of time between the end of tunneling and the microwave spin rotation is $>1$ ms.  Second, the magnetic quantization field provides a large energy difference between the $\ket{\uparrow,\uparrow}$ and $\ket{\downarrow,\downarrow}$ spin states, which would result in parity oscillations at a significantly faster rate than the primary oscillations observed in Fig.~3(b).  It is important to note that these single-spin coherences and energy offsets do not effect the interference of the two atoms; we only note them because the states that could mimic the signal shown in Fig.~3(b) require large single-spin coherences to be present throughout the experiment.  
For our operation, it is the spatial coherence of the atoms that matters, which can be seen to be much longer than any of the relevant timescales in Fig.~2.

The elimination of $\rho_{\uparrow\uparrow,\downarrow\downarrow}$ and $\rho_{\downarrow\downarrow,\uparrow\uparrow}$ terms is also important for the entanglement bound quoted in the main text, because we need to compare the magnitude of $\rho_{\uparrow\downarrow,\downarrow\uparrow}$ to the bound for entanglement, and we can only safely relate the measured parity to $\rho_{\uparrow\downarrow,\downarrow\uparrow}$ in the absence of additional coherences. With this assumption, we can compare the measured parity oscillation contrast to the bound on the amplitude of the oscillations $\pm A$, as shown in Fig.~3, which is derived by assuming the state is described by a separable density matrix composed of the single-particle density matrices $\rho^{1}$ and $\rho^{2}$ representing particle 1 and particle 2, respectively. Then, the magnitude of the parity would not exceed $A= 2 \,|\rho^{1}_{\uparrow,\downarrow}||\rho^{2}_{\downarrow,\uparrow}|\leq 2\, \big(\rho^{1}_{\uparrow,\uparrow}\rho^{1}_{\downarrow,\downarrow}\big)^{1/2}\big(\rho^{2}_{\uparrow,\uparrow}\rho^{2}_{\downarrow,\downarrow}\big)^{1/2}=2\, \big(\rho_{\uparrow\uparrow,\uparrow\uparrow}\;\rho_{\downarrow\downarrow,\downarrow\downarrow}\big)^{1/2}$~\cite{kaufman_entangling_2015}.

It is also interesting to calculate the fidelity in the worst case scenario in which the $\rho_{\uparrow\uparrow,\downarrow\downarrow}$ term maximally contaminates the parity measurement.  In general, the parity contrast is given by $C_\Pi=2({\rm Re}(\rho_{\uparrow\downarrow,\downarrow\uparrow})-{\rm Re}(\rho_{\uparrow\uparrow,\downarrow\downarrow})\cos(2 \phi)+{\rm Im}(\rho_{\uparrow\uparrow,\downarrow\downarrow})\sin(2 \phi))$, where $\phi$ denotes the phase of the microwave drive with respect to the phase of the spin.  We consider the three following detrimental effects: 1) The erroneous preparation in the $\ket{\uparrow,\uparrow}$ and $\ket{\downarrow,\downarrow}$ states happens to create maximum amplitude of $\rho_{\uparrow\uparrow,\downarrow\downarrow}$. 2) The phase coherence between the $\ket{\uparrow,\uparrow}$ and $\ket{\downarrow,\downarrow}$ state does not decay quickly before the measurement.  3) The microwave phase $\phi$ coincides with the complex phase of $\rho_{\uparrow\uparrow,\downarrow\downarrow}$.  If all three of these cases were to conspire to maximally contaminate the parity, we would find
	\begin{align*} 
    \label{eq:Pi3}
		C_\Pi&=2({\rm Re}(\rho_{\uparrow\downarrow,\downarrow\uparrow}) - \\
		&~~~~\left( \rm max\left\{{\rm Re}(\rho_{\uparrow\uparrow,\downarrow\downarrow})\cos(2 \phi)-{\rm Im}(\rho_{\uparrow\uparrow,\downarrow\downarrow})\sin(2 \phi)\right\}\right) )\\
		&=2({\rm Re}(\rho_{\uparrow\downarrow,\downarrow\uparrow})-\rm abs(\rho_{\uparrow\uparrow,\downarrow\downarrow}))\\
		&\geq2({\rm Re}(\rho_{\uparrow\downarrow,\downarrow\uparrow})-\sqrt{\rho_{\uparrow\uparrow,\uparrow\uparrow}\rho_{\downarrow\downarrow,\downarrow\downarrow}}).
	\end{align*}
Thus in the worst case we have 
	\begin{align}
    	{\rm Re}(\rho_{\uparrow\downarrow,\downarrow\uparrow})_{\rm worst\,case}&=C_\Pi/2+\sqrt{\rho_{\uparrow\uparrow,\uparrow\uparrow}\rho_{\downarrow\downarrow,\downarrow\downarrow}},
	\end{align}
with $\rho_{\uparrow\uparrow,\uparrow\uparrow}=\left(0.12\pm0.02\right)$ and $\rho_{\downarrow\downarrow,\downarrow\downarrow}=\left(0.01\pm0.01\right)$, this gives a fidelity $\mathcal{F}_{S_0,\text{worst-case}}=\left(0.58\pm0.03\right)$.

\subsection{$\chi^2$ analysis of fit to parity oscillation}
For the data in Fig. 2(b), we fit the posts	elected data and obtain $\chi^2=\sum_i{\left(\frac{y_{{\rm fit},i}-\bar{y_i}}{\delta_i}\right)^2}$, where $\bar{y_i}$ is the mean measured parity at the $i^{\rm th}$ data point, $\delta_i$ is the standard deviation for each data point, and $y_{{\rm fit},i}$ is the fitted value. We obtain $\chi^2=16.24$. The fit involved 15 data points and 4 parameters leaving 11 degrees of freedom.  Thus, the reduced $\chi^{2}_{11} = 1.48$ (p-value 0.13), which we consider an acceptable quality of fit. 

\subsection{Nonparametric bootstrap analysis of parity data}
To check the consistency of our data, we use nonparametric bootstrap resampling~\cite{Efron1993}. For each data point in Fig.~3 we randomly select 1000 measurements out of the 1000 trials with replacement, to obtain a resampled data set. Then we apply the same methods for data analysis in the main text and obtain the fitted amplitudes for the parity oscillation. We repeat this resampling process 5000 times, and we obtain a histogram of the fitted amplitudes where the mean and the deviation give the resampled fit amplitude and uncertainty. These resampled fit parameters reproduce the values obtained from the original data from the non-linear least-square fit. 

\begin{figure}[t!]
	\begin{center}
		\includegraphics[width=0.9\columnwidth]{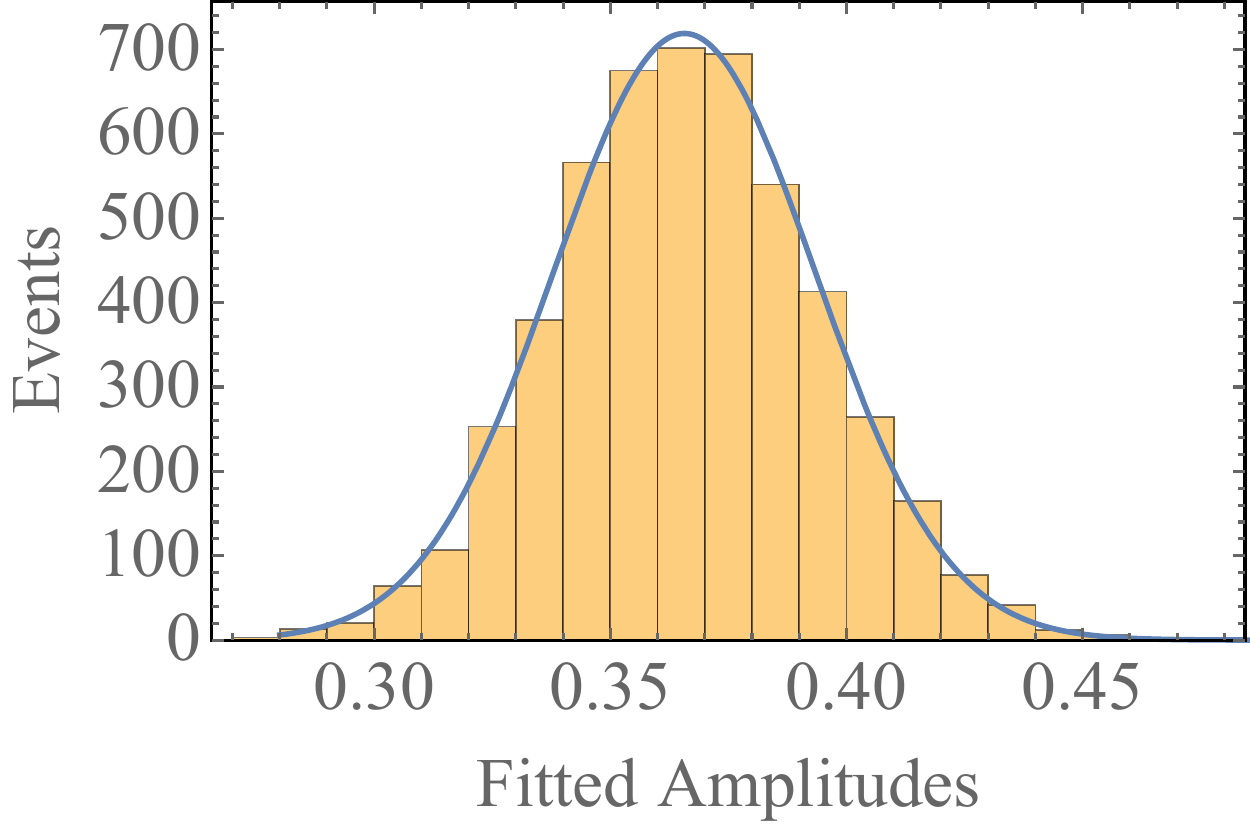}
		\caption{
			A histogram of the fitted amplitude for resampled data in Fig.~3(b). Here a Gaussian fit to the histogram gives the mean of the resampled fitted amplitudes of 0.366 with $\sigma$=0.039. This matches the fitted amplitudes $\left(0.36\pm0.03\right)$ of the original data set. 
		}
		\label{bootstrapping}
	\end{center}
\end{figure} 

\subsection{Known contributions to finite fidelity}

While the analysis above fully characterizes the synthesized state, it is useful to estimate the contributions of known errors to the infidelity of $\ket{S_0}$ state creation.  Because we filter out errors due to atom loss, this type of error does not contribute to reduced contrast of our parity oscillations, but we remain sensitive to other imperfections, discussed below, that affect the indistinguishability of the bosonic particles in degrees of freedom other than spin.

The effect of ground-state preparation and finite tunneling contrast (as see in Fig.~2) can be estimated based on measurements of the two-particle quantum interference contrast $C_\text{HOM}\simeq0.6$, which is similarly reduced by the loss of indistinguishability. The contribution of the spin preparation efficiency can be separately characterized by directly measuring the spin populations after the preparation sequence. In this case, we find the dominant error to be residual population in the $\ket{\uparrow,\uparrow}$ state. In the case of perfect HOM contrast this state would be post-selected out when they end on the same well.  However, there is an error contributed when the atoms end up in separate wells, and hence this spin-state contributes an additional error to the parity oscillation contrast proportional to $\left(1-C_\text{HOM}\right)$.  Then, taking into account the filtering process of other spin states, we expect a parity of $\Pi=\frac{(1-\rho_{\uparrow\uparrow,\uparrow\uparrow})C_{\rm HOM}}{-2+C_{\rm HOM}+C_{\rm HOM}\rho_{\uparrow\uparrow,\uparrow\uparrow}}$, which is approximately -0.4 given that $\rho_{\uparrow\uparrow,\uparrow\uparrow}=0.12$ and $C_{\rm HOM}$=0.6. This expected contrast is consistent with our measurements, thus we expect that improving the state preparation and tunneling procedures, which improves the indistinguishability of the atoms, the singlet state fidelity should also be improved.

\end{document}